\documentstyle[emulateapj,psfig]{article}

\begin{document}
\submitted{Astrophysical Journal (2000), in press}
\title{Radio Sources in Low-Luminosity Active Galactic Nuclei. 
II. VLBI Detections of Compact Radio Cores and Jets in a Sample of LINERs}

\author{Heino Falcke}
\affil{Max-Planck-Institut f\"ur Radioastronomie, Auf dem H\"ugel 69,
53121 Bonn, Germany \\(hfalcke@mpifr-bonn.mpg.de)}

\author{Neil M. Nagar \and  Andrew S. Wilson\altaffilmark{1}}
\affil{Astronomy Department, University of Maryland, College Park,
MD 20742-2421 \\(neil,wilson@astro.umd.edu)}
\altaffiltext{1}{Adjunct Astronomer, Space Telescope Science Institute}

\author{James S. Ulvestad}
\affil{National Radio Astronomy Observatory, P.O. Box 0, Socorro, NM 87801\\(julvesta@nrao.edu)}

\begin{abstract}
We have used the VLBA at 5 GHz to observe all galaxies with nuclear
radio flux densities above 3.5 mJy found in a VLA survey at 15 GHz of a sample
of nearby LINER galaxies. All galaxies were detected revealing high
brightness temperature ($T_{b} \ga 10^8$ K) radio sources.  Free-free
emission is unlikely since it greatly overpredicts the soft X-ray
luminosities. We infer the presence of AGN-like, non-thermal radio
emission most likely powered by under-fed black holes.  Together with
our VLA sample we estimate from our observations that at least half of
LINER galaxies host genuine AGN.  We find no evidence for highly
inverted radio cores as predicted in the ADAF model: the
(non-simultaneous) spectral indices are on average around
$\alpha=0.0$.  In the two brightest sources we detect some extended
emission, which appears to originate in jets in at least one of these
galaxies. Together with the spectral indices this suggests that the
nuclear emission at centimeter radio waves is largely dominated by
emission from radio jets, very similar to the situation in more
luminous AGN. The energy released in these jets could be a significant
fraction of the energy budget in the accretion flow.

\end{abstract}

\keywords{galaxies: active --- galaxies: jets ---
galaxies: nuclei --- galaxies: structure --- galaxies: Seyferts --- radio
continuum: galaxies}

\section{Introduction}
The evidence for supermassive black holes in the nuclei of most
galaxies has become much stronger recently. Some of the best cases are
the Milky Way (Eckart \& Genzel 1997), NGC 4258 (Miyoshi et al. 1995),
and a number of other nearby galaxies (Richstone et al. 1998) where
convincing dynamical evidence for black holes exists. In quasars and
radio galaxies their existence is commonly inferred from the huge
energy output of the active galactic nucleus (AGN) which is probably
powered by accretion onto the black hole.  However, despite the
alleged presence of black holes in both cases, there is a huge span in
luminosity between weakly active galaxies like the Milky Way and
AGN. The question of how these central engines are related to each
other and why they appear so different despite being powered by the
same type of object is therefore of major interest. For many nearby
galaxies with low luminosity nuclear emission lines, it is not even
clear whether they are powered by an AGN or by star formation. This is
especially true for Low Ionization Nuclear Emission Region (LINER)
galaxies (Heckman 1980), some of which can be explained in terms of
aging starbursts (Alonso-Herrero et al.~2000).

One of the best ways to probe the very inner parts of these engines is
to study the compact radio sources found in many AGN. Indeed, despite
their low optical luminosity, quite a few nearby galaxies have such
radio sources in their nuclei (e.g. Jones et al.~1981), prominent
cases in spiral galaxies being the Milky Way (Sgr A*) and M~81 (see
Bietenholz et al. 1996). In addition some relatively nearby elliptical
galaxies such as M87 and NGC~1275 appear as low-power FR\,I radio
galaxies and also contain well-known compact radio cores (Cohen et
al.~1969; Schilizzi et al.~1975).

These radio sources resemble the cores of radio-loud quasars, showing
a very high brightness temperature and a flat to inverted radio
spectrum that extends up to sub-mm wavelengths. Models proposed for
these low luminosity radio nuclei are either a scaled AGN model, in
which the core is the synchrotron self-absorbed base of a radio jet
coupled to an underluminous accretion disk (Falcke, Mannheim, \&
Biermann 1993; Falcke 1996; Falcke \& Biermann 1999) or an
advection-dominated accretion flow (ADAF; Narayan et al.~1998; see
also Melia 1992; Fabian \& Rees 1995).

Earlier surveys have shown that E and S0 galaxies often have compact,
flat-spectrum radio sources in their nuclei (Wrobel \& Heeschen 1984,
1991; Sadler et al.~1989; Slee et al.~1994).  Some of the most
prominent flat-spectrum nuclear radio sources in nearby galaxies are
found in galaxies with LINER nuclear spectra (O'Connell \& Dressel
1978), but so far there has been no comprehensive study of radio
nuclei in a significant sample of LINER galaxies, which make up the
majority of galaxies with low-level nuclear activity. We have,
therefore, recently conducted a survey of LINER galaxies with the Very
Large Array (VLA; Thompson et al. 1980) in its A configuration at 15
GHz (resolution $\sim$0\farcs15) to search for compact radio emission
(Nagar et al.~2000, Paper I). The sources were drawn from the
extensive and sensitive spectroscopic study of a complete,
magnitude-limited sample of 486 nearby galaxies (Ho, Filippenko, \&
Sargent 1995), one third of which showed LINER-like activity (Ho,
Filippenko, \& Sargent 1997). From those active galaxies with a LINER
spectrum a subsample of 48 bright sources was drawn with no
well-defined selection criterion other than that they had been
observed with other telescopes as well, e.g. ROSAT, the HST (UV
imaging, Maoz et al. 1996, Barth et al. 1998), and the VLA at 15 GHz
in A configuration (Nagar et al. 2000) and at 1.4 and 8.4 GHz in A and
B configuration (van Dyk \& Ho 1997).  The sample also included so
called transition objects which have spectra intermediate between
LINER and \ion{H}{2} region galaxies. While the project was being
conducted a few sources in the original sample were re-classified as
low-luminosity Seyfert galaxies. However, only one out of the ten
sources discussed here has a ``pure'' Seyfert spectrum.

The 15~GHz VLA survey found a surprisingly large number (15 out of 48)
of galaxies with compact radio cores and flat spectral indices.  Here
we present Very Long Baseline Array (VLBA; Napier et al. 1994)
observations of the eleven brightest of these galaxies to investigate
the central region of LINER galaxies at the sub-parsec scale and
clarify the nature of their radio cores.

\section{Sample selection and Observations}

{}From our A configuration VLA survey (Nagar et al.~2000), we selected
all eleven galaxies with both nuclear flux densities above 3.5 mJy at
15 GHz and a flat spectrum ($\alpha>-0.5,\;
S_\nu\propto\nu^\alpha$). Only one source, NGC~2655, was above the
flux density limit and was excluded because of its steep spectrum. The
flux density limit was chosen so that we could detect all sources with
the VLBA in snapshot mode in a single 12 hr observation if most of the
15 GHz emission were indeed compact on milli-arcsecond (mas) scales.

The observations were performed on 1997 June 16 with all ten antennas
of the VLBA at 4.975 GHz.  Because the sources are so faint we had to
use phase-referencing. Observation of each program source was
preceded by observation of a bright nearby phase calibrator. The
observation sequence was cal-source-cal-source, which was repeated for
three different hour angles to improve (u,v)-coverage. The integration
times per scan were roughly 1-2 minutes for the calibrators and
$\sim$8 minutes for the program galaxies. The total integration time per
galaxy was therefore about 45 minutes.

The data for the calibrators were fringe-fitted, imaged, and
self-calibrated in AIPS and the solutions for antenna gains, phases,
delays, and clock rates were transferred to the program sources. The
data for the program sources were then imaged and cleaned. Since the
initial coordinates were taken from our VLA observations we had in
most cases a position uncertainty of $\sim$0\farcs05$-$0\farcs1 and
hence we first imaged a wide field of view (0\farcs4). The map center
was then shifted to the highest peak in this image and the source was
re-mapped and phase self-calibrated. A final amplitude and phase
self-calibration was also performed for a long solution interval (25
minutes) and the sources were imaged.

The observations of NGC~3147 failed since the phase-calibrator was not
detected, thus reducing our sample to ten sources. We note, however,
that this source, together with NGC~2655 and NGC~4143, which is just
barely under our flux density limit, are included in further VLBA
observations (Nagar et al., in preparation).

\section{Results}
We detected all 10 sources with the VLBA.  The results are shown in
Table 1, where we list the galaxy names, distances (from Ho et
al.~1997), Hubble galaxy type from RC3 (de Vaucouleurs et al. 1991)
and spectroscopic classification from Ho et al.~(1997) in columns 1 -
4.  The sources are equally distributed between early- and late-type
galaxies. We also give the positions of the radio sources (columns 5
and 6). The internal errors of the positions should be about the beam
size (2.5 mas) or better for the stronger ones, but the absolute
astrometric accuracy is limited by the uncertainty in the positions of
the phase calibrators, (listed in column 7). In addition, we list the
total and peak flux densities in the VLBA maps and the brightness
temperature (column 10, defined below).  Our $1\sigma$ rms noise is
typically 0.2 mJy and hence for the weakest source we obtain a dynamic
range of 15:1. For extended sources we give the position angle (column
11) of an elliptical Gaussian component fitted to the core. For
comparison we also list the peak VLA flux density at 15 GHz (column
12) from our previous survey and the (non-simultaneous) spectral index
$\alpha$ between the peak VLA 15 GHz and total VLBA 5 GHz flux
densities (column 13).

\begin{deluxetable}{lrrcllrrrrrrr}
\scriptsize
\tablecaption{Properties of VLBA LINER sample}
\tablehead{
\colhead{(1)}&\colhead{(2)}&\colhead{(3)}&\colhead{(4)}&\colhead{(5)}&\colhead{(6)}&\colhead{(7)}&\colhead{(8)} & \colhead{(9)}&\colhead{(10)}&\colhead{(11)}&\colhead{(12)}&\colhead{(13)}\\
\colhead{Name}&\colhead{D}&
\colhead{T}&\colhead{spec.}&\colhead{R.A.}&\colhead{Dec.}&\colhead{$\Delta_{\rm 
pos}$}&\colhead{$S_{\rm 5 GHz}^{\rm total}$}&\colhead{$S_{\rm 5
GHz}^{\rm peak}$}&\colhead{$T_{\rm b}$}
& P.A.&\colhead{$S_{\rm 15 GHz}^{\rm VLA}$}&\colhead{$\alpha$}\\
\colhead{}&\colhead{[Mpc]}&\colhead{}&\colhead{}&\colhead{(J2000)}&\colhead{(J2000)}&\colhead{[mas]}&\colhead{[mJy]}&\colhead{[mJy]}&\colhead{[$10^8$K]}&\colhead{[$^\circ$]}&\colhead{[mJy]}&\colhead{}}
\tablecolumns{13}
\startdata
NGC~266  &62.4& 2&L1.9& 00 49 47.8174  &+32 16 39.749& 55 &3.8  &3.2  & 0.25 &    & 4.1  & 0.07 \\
NGC~2787 &13.0&-1&L1.9& 09 19 18.6095  &+69 12 11.690& 10 &11.5 &11.2 & 0.87 &    & 7.0  &-0.45\\
NGC~3169 &19.7& 1&L2.0& 10 14 15.0500  &+03 27 57.844& 14 &6.6  &6.2  & 0.48 &    & 6.8  & 0.03 \\
NGC~3226 &23.4&-5&L1.9& 10 23 27.0113  &+19 53 54.496& 150&4.8  &3.5  & 0.27 &(64)& 5.0  & 0.04\\
NGC~4203 &9.7 &-3&L1.9& 12 15 05.0519  &+33 11 50.359& 55 &8.9  &8.9  & 0.69 &    & 9.5  & 0.06\\
NGC~4278 &9.7 &-5&L1.9& 12 20 06.8254  &+29 16 50.715& 1  &87.3 &37.2 & 2.9  &163 & 88.3 & 0.01\\%
NGC~4565 &9.7 & 3&S1.9& 12 36 20.7820  &+25 59 15.632& 55 &3.1  &3.2  & 0.25 &    & 3.7  & 0.16\\
NGC~4579 &16.8& 3&L1.9& 12 37 43.5222  &+11 49 05.488& 1  &21.3 &21.3 & 1.7  &    & 27.6 & 0.23\\%
NGC~5866 &15.3&-1&T2.0& 15 06 29.4989  &+55 45 47.568& 1  &8.4  &7.0  & 0.55 &(11)& 7.1  &-0.15\\
NGC~6500 &39.7& 2&L2.0& 17 55 59.7827  &+18 20 17.661& 14 &83.6 &35.8 & 2.8  & 39 & 83.5 & 0.00\\%
\enddata

\tablecomments{The columns are: (1) galaxy name; (2) distance in Mpc; 
(3) host galaxy morphological type from RC3; 
(4) spectroscopical AGN classification:
L=LINER, S=Seyfert, T=LINER -- \ion{H}{2} region transition galaxy; (5~\&~6) VLBI J2000 coordinates; 
(7) position uncertainty of phase referencing calibrator in
milli-arcsecond (mas);
(8) total flux
density at 5~GHz (VLBA); (9) peak flux density at 5~GHz (VLBA), the
statistical 1$\sigma$ error is 0.2 mJy; 
(10) brightness temperature in $10^8$ K;
(11)
position angle, measured N through E, of 5~GHz radio core if extended; 
values in brackets are very uncertain;
(12) peak VLA flux density at 15~GHz (Nagar et al. 2000); 
(13) spectral index between peak 15~GHz (VLA) and total 5~GHz (VLBA) flux
     densities. 
Columns 2-4 are from Ho et al.~(1997).}

\end{deluxetable}
\begin{figure*}\label{radiomap}
\centerline{\psfig{figure=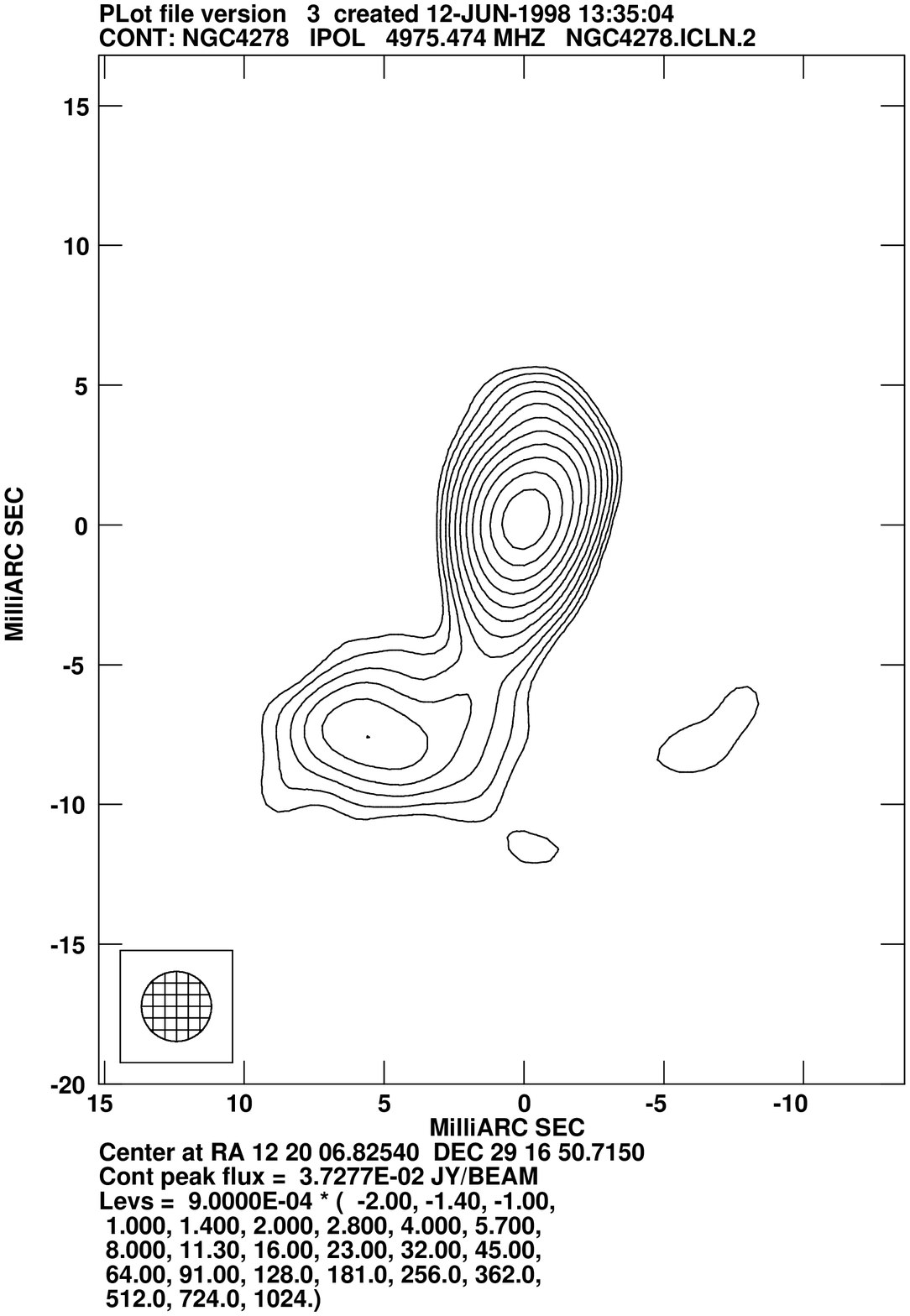,height=0.35\textwidth,bbllx=2.1cm,bburx=19.2cm,bblly=4.9cm,bbury=25.4cm,clip=}\quad\psfig{figure=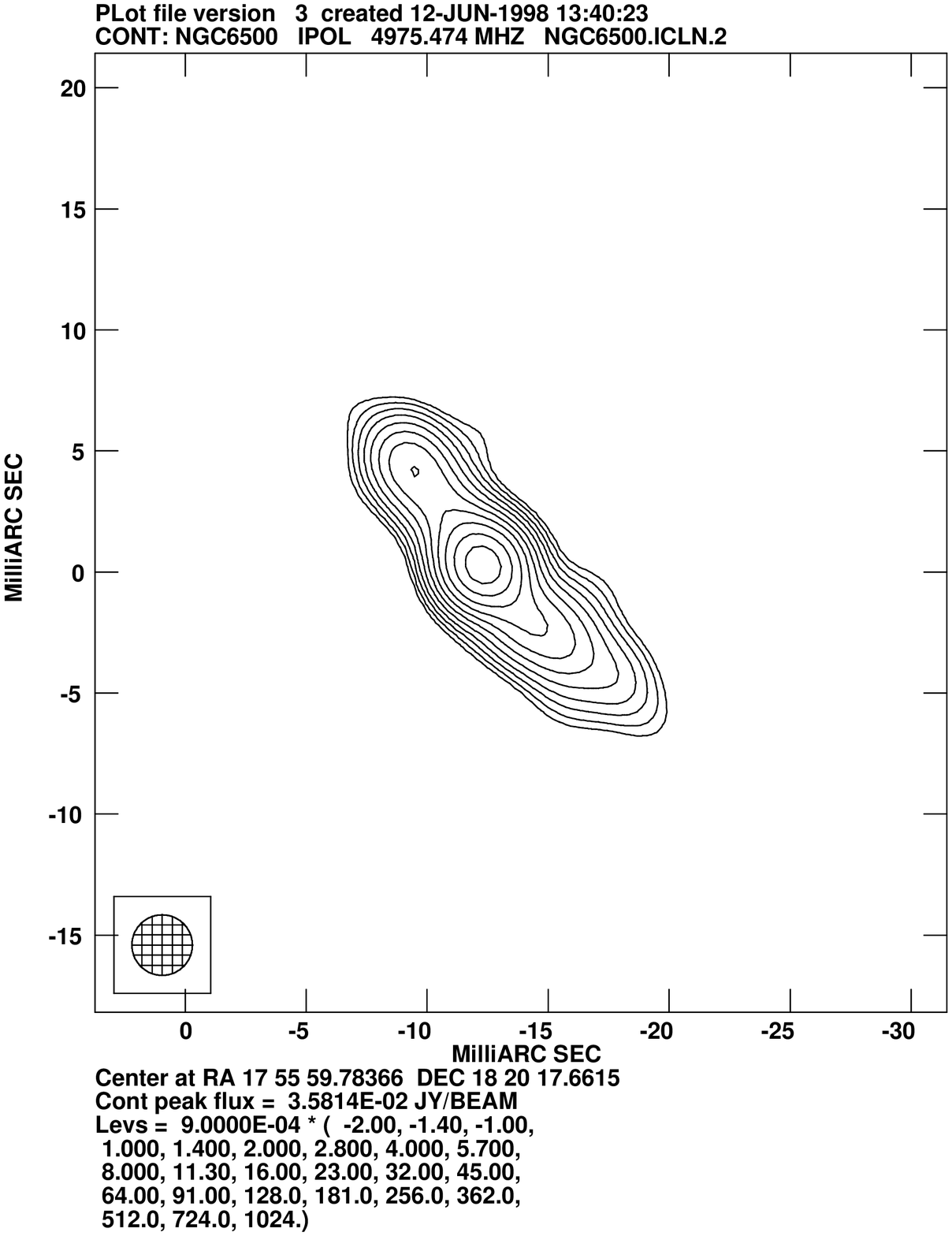,height=0.35\textwidth,bbllx=1.425cm,bburx=19.95cm,bblly=5.2cm,bbury=25.05cm,clip=}}

\caption[]{VLBA maps of NGC~4278 (left) and NGC~6500 (right). The beam is 2.5 mas and contours are integer powers
of $\sqrt{2}$, multiplied by the $\sim5\,\sigma$ noise level of 0.9
mJy.  The peak flux densities are 37.3 mJy and 35.8 mJy respectively.}
\end{figure*}

The two brightest sources in our sample, NGC~4278 \& NGC~6500, for
which we have the largest dynamic range, show core plus jet structures
(Fig.~1). While the compact emission in these two sources was known
before (Jones et al.~1981), the new observations now show the extended
emission with unprecedented quality. NGC~6500 has a core and a
symmetric two-sided jet, while NGC~4278 has an extended core and an
elongated source towards the SE. The other sources are point-like with
the possible exceptions of NGC~3226 and NGC~5866, although phase
errors may be responsible for the extension in these faint sources.
The spectral indices range from $\alpha=-0.5$ to $\alpha=0.2$, with an
average $\left<\alpha\right>=0.0\pm0.2$. We note that the VLBA does
not provide spacings as short as the VLA. The VLBA measurements at 5
GHz may then underestimate the true flux density if the sources are
extended below the 150 mas scale, so the actual spectrum might be less
inverted (i.e. the spectral index smaller) than we measure here. The
statistical error in our flux density measurements produces a 1
$\sigma$ error of up to $\pm$0.1 in the 5-to-15 GHz spectral index for
the weakest sources.

{}From the flux densities and the sizes we have measured, we can
calculate the brightness temperatures for a Gaussian flux density distribution
with the following equation


\begin{equation}
T_{\rm b}=7.8\times10^6\;{\rm K} \left({S_{\nu}\over{\rm mJy}}\right)\left({\theta\over{\rm 2.5\,mas}}\right)^{-2}\left({\nu\over{\rm 5\,GHz}}\right)^{-2}
\end{equation}
where $\theta$ is the FWHM of the Gaussian beam (e.g. Condon et al. 1982). Using our beam
size of 2.5 mas and the peak 5 GHz flux densities in Table 1, we find
brightness temperatures in the range $T=0.25-2.9\times10^8$ K for our
sample, with an average brightness temperature for all sources of
$\left<T_{\rm b}\right>=1.0\times10^8$ K. Since most of our sources
are unresolved, these values are usually lower limits.


\section{Discussion}

Our result has a number of interesting implications. The presence of
high brightness temperature radio cores in our LINER sample confirms
the presence of AGN-like activity in these galaxies. It is unlikely
that the radio sources represent free-free emission, as has been
claimed for example in NGC~1068 (Gallimore, Baum, O'Dea 1997), since a
much higher soft X-ray luminosity than is typically observed in
low-luminosity AGN would result.  The emission coefficient for thermal
bremsstrahlung from a gas at temperature $T$ is (e.g. Longair 1992,
eq.~3.43)
\begin{eqnarray}
\epsilon_\nu~&=& 6.8 \times 10^{-51} Z^2 T^{{\small{-\onehalf}}} N_p N_e\nonumber\\
&&\cdot	      g(\nu, T) exp(- h\nu/kT)\; {\rm W m^{-3} Hz}^{-1}
\end{eqnarray}
where $g(\nu, T)$ is the Gaunt factor.  If we consider a plasma at
temperature $T\simeq10^8$ K, then at 5~GHz, the exponential factor in
eqn. (2) is 1, while the Gaunt factor is $\sim$12.  In the soft X-ray
regime, taken as 0.4~keV to 2~keV, the Gaunt factor varies between 1.7
and 0.8, respectively, while the exponential factor in eq. (2) varies
between 0.95 and 0.8, respectively.  Therefore, the luminosity, per
Hertz, at 0.4~keV and 2~keV is $\sim$0.25 times and $\sim$0.05 times
that at 5~GHz, respectively.  The geometric mean monochromatic
luminosities of the nuclei observed by us is 10$^{27.5\pm0.6}$ erg
sec$^{-1}$ Hz$^{-1}$ at 5~GHz. If this emission traces thermal
bremsstrahlung we would expect the total 0.4--2~keV luminosity of
these nuclei to be 10$^{43.9}$ erg s$^{-1}$.  However, the observed
0.4-2~keV luminosities for low-luminosity AGN tend to be of the order
of 10$^{39-40}$ erg s$^{-1}$ (e.g. Ptak et al. 1999)---many orders of
magnitude lower and thus rendering a thermal origin of the radio
emission very unlikely. Of course photoelectric absorption could
attenuate some of the soft X-ray emission, however, since seven of our
galaxies show broad H$\alpha$ emission (spectral type 1.9) the
absorption should only be moderate. To make this argument more
watertight one will need to investigate multiwavelength data for our
galaxies on a case-by-case basis.

On the other hand, the compact, flat-spectrum cores we have found are
similar to those typically produced in many AGN. Hence we can take the
presence of compact, non-thermal radio emission as good evidence for
the presence of an AGN in our galaxies.  The 100\% detection rate with
the VLBA, based on our selection of flat-spectrum cores found in a 15
GHz VLA survey, shows that for statistical purposes we could have
relied on the VLA alone for identification of these compact, high
brightness radio sources.  Hence, with 15 GHz VLA surveys of nearby
galaxies one has an efficient tool for identifying low-luminosity
AGN. This complements other methods for identifying AGN, such as
searching for broad emission-lines or hard X-rays, and has the
advantage of not being affected by obscuration.

If we only consider galaxies with a LINER spectrum, we found at least
eleven flat-spectrum radio cores at 15 GHz in a sub-sample of 24
LINERs observed by Nagar et al.~(2000).  Eight of these eleven LINERs
are included in our sample here, yielding a lower limit to the AGN
fraction for LINERs of at least $33\pm12$\% (8/24). Based on our 100\%
detection rate of these flat-spectrum cores with the VLBA, we can,
however, argue that all eleven flat spectrum sources found in the VLA
study are likely to be AGN, raising the AGN fraction of LINERs to at
least $46\pm14$\% (11/24). These ratios do not change significantly if
we include the galaxies classified as Seyferts. Since the selection of
our parent sample is not very well defined, we could still be subject
to an unquantifiable bias. This can be minimized by studying the
radio emission of a distance-limited sample, which we plan in a future
paper. First results (Falcke et al.~1999) seem to indicate that the
bias is not large.

The two brightest radio sources in our sample show extended structure
suggestive of jet-like outflows, and the other seven sources are
unresolved or slightly resolved.  Our very limited dynamic range is
not good enough to prove or exclude the presence of jets for the
latter. Moreover, VLBA observations of M81 (Bietenholz et al.~2000)
have shown that jets in low-luminosity AGN can be very compact and
difficult to detect. The only clue we therefore have is the spectrum
which is flat or slightly inverted. Such a spectrum is obtained in jet
models (Blandford \& K\"onigl 1979; Falcke 1996; Falcke \& Biermann
1999), where the spectral index ranges from $\alpha=0.0$ to 0.23 as a
function of inclination angle to the line-of-sight. In no case do we
find a spectral index as high as $\alpha=0.4$ as predicted in the ADAF
model (Yi \& Boughn 1998). This does not necessarily exclude the ADAF
model, but argues for the parsec scale radio emission at centimeter
radio waves being dominated by another component, such as a radio jet
or a wind. A combination of an underluminous disk or an ADAF and a
radio jet is one possibility (e.g.~Donea, Falcke, \& Biermann 1999).

Assuming the cores are produced by randomly oriented, maximally
efficient jets from supermassive black holes (of order $10^8 M_\odot$)
we can use eq. (20) of Falcke \& Biermann (1999) to calculate that for
an average monochromatic luminosity of $10^{27.5}$ erg sec$^{-1}$
Hz$^{-1}$ at 5 GHz the jets would require a minimum {\em total} jet
power of order $Q_{\rm jet}\ga10^{42.5}$ erg sec$^{-1}$. Compared to
quasars this is a rather low value and supports the conclusion, based
on their low UV and emission line luminosities, that the cores are
powered by under-fed black holes. On the other hand this jet power is
well within the range of the bolometric luminosity of typical
low-luminosity AGN ($10^{41-43}$ erg sec$^{-1}$; Ho 1999) and,
compared to radiation, jets could be a significant energy loss channel
for the accretion flow.

\acknowledgements This research was supported by NASA and NSF under grants
NAG8-1027 and AST9527289, respectively.  HF is supported by Fa DFG
grant 358/1-1\&2.  The National Radio Astronomy Observatory is a
facility of the National Science Foundation, operated under
cooperative agreement by Associated Universities, Inc.

\clearpage

\onecolumn


\begin{thebibliography}{xxx}
\bibitem[]{}Alonso-Herrero, A., Rieke, M. J., Rieke, G. H., \&
Shields, J. C. 2000, ApJ 530, 688
\bibitem[]{}Barth, A. J., Ho, L. C., Filippenko, A. V., \& Sargent, W. L. W.  1998, \apj, 496, 133
\bibitem[Bietenholz et al. 1996]{1996ApJ...457..604B} Bietenholz, M. F., et al. 1996, \apj, 457, 604 
\bibitem[Bietenholz et al. 2000]{2000ApJ...457..604B} Bietenholz, M.F., Bartel, N., Rupen, M.P. 2000, \apj, 532, 895
\bibitem[Blandford \& K\"onigl (1979)]{1979ApJ...232...34B} Blandford, R. D. \& 
K\"onigl, A. 1979, \apj, 232, 34 
\bibitem[Cohen, et al. (1969)]{1969ApJ...158L..83C} Cohen, M. H., Moffet, 
A. T., Shaffer, D., Clark, B. G., Kellermann, K. I., Jauncey, D. L. \& 
Gulkis, S. 1969, \apjl, 158, L83 
\bibitem[Condon Condon Gisler \& Puschell 1982]{1982ApJ...252..102C} 
Condon, J. J., Condon, M. A., Gisler, G. \& Puschell, J. J. 1982, \apj, 
252, 102 
\bibitem[]{}de Vaucouleurs, G., de Vaucouleurs, A., Corwin, H. G., Jr., Buta,
R. J., Paturel, G., \& Fouqu\'e, R. 1991, Third Reference Catalogue of
Bright Galaxies (New York: Springer)
\bibitem[]{}{Donea, A. C., Falcke, H., Biermann, P. L. 1999, in: ``The Central Parsecs of the Galaxy'',
eds. H. Falcke, A. Cotera, W. Duschl, F. Melia, M. Rieke, ASP Conf. Series, Vol. 186, p. 162}
\bibitem[Eckart \& Genzel 1997]{1997MNRAS.284..576E} Eckart, A., \& Genzel, 
R. 1997, \mnras, 284, 576 
\bibitem[Fabian \& Rees 1995]{1995MNRAS.277L..55F} Fabian, A. C., \& Rees, 
M. J. 1995, \mnras, 277, L55 
\bibitem[]{}{Falcke, H. 1996, ApJ, 464, L67}
\bibitem[]{}{Falcke, H., \& Biermann, P. L. 1996, A\&A, 308, 321}
\bibitem[]{}{Falcke, H., \& Biermann, P. L. 1999, A\&A, 342, 49}
\bibitem[]{}{Falcke, H., Mannheim, K., \& Biermann, P. L. 1993, A\&A, 278, L1}
\bibitem[]{}{Falcke, H., Nagar, N. M., Wilson, A. S., Ho, L. C.,
Ulvestad, J. S. 1999, in: ``Black Holes in Binaries and Galactic
Nuclei'', ESO workshop, eds. L. Kaper, E. P. J. van den Heuvel,
P. A. Woudt, Springer Verlag, in press}
\bibitem[Gallimore Baum \& O'Dea 1997]{1997Natur.388..852G} Gallimore, J. 
F., Baum, S. A. \& O'Dea, C. P. 1997, \nat, 388, 852 
\bibitem[]{}{Heckman, T. M. 1980, A\&A, 87, 152}
\bibitem[Ho 1999]{1999ApJ...516..672H} Ho, L. C. 1999, \apj, 516, 672 
\bibitem[]{}{Ho, L. C., Filippenko, A. V., \& Sargent, W. L. W. 1995, ApJS, 98, 477}
\bibitem[]{}{Ho, L.~C., Filippenko, A.~V., \& Sargent, W.~L.~W. 1997, ApJS, 112, 315}
\bibitem[Jones, Terzian \& Sramek (1981)]{1981ApJ...246...28J} Jones, D. L., 
Terzian, Y. \& Sramek, R. A. 1981, \apj, 246, 28 
\bibitem[]{}Longair, M. 1992, High Energy Astrophysics Vol. I,
2nd edition (Cambridge: Cambridge University Press)
\bibitem[Maoz et al. 1996]{1996ApJS..107..215M} Maoz, D. , Filippenko, A. 
V., Ho, L. C., Macchetto, F. D., Rix, H. -W.  \& Schneider, D. P. 1996, 
\apjs, 107, 215 
\bibitem[]{}{Melia, F. 1992, ApJ, 398, L95}
\bibitem[Miyoshi et al. 1995]{1995Natur.373..127M} Miyoshi, M., Moran, J., 
Herrnstein, J., Greenhill, L., Nakai, N., Diamond, P. \& Inoue, M. 1995, 
\nat, 373, 127 
\bibitem[]{}{Nagar, N. M, Falcke, H., Wilson, A. S., \& Ho,
L. C. 2000, ApJ, in press}
\bibitem[]{}{Napier, P. J., Bagri, D. S., Clark, B. G., Rogers, A. E. E.,
Romney, J. D., Thompson, A. R., \& Walker, R. C. 1994, Proc.
IEEE, 82, 658}
\bibitem[]{}{Narayan, R., Mahadevan, R., Grindlay,  J.~E., Popham, R. G., \& Gammie, C. ~1998, ApJ, 492, 554}
\bibitem[]{}{O'Connell, R. W., Dressel, L. L. 1978, Nature, 276, 374}
\bibitem[]{}{Ptak, A., Serlemitsos, P., Yaqoob, T., \& Mushotzky, R.
	     1999, \apjs, 120, 179}
\bibitem[]{}{Richstone, D., et al. 1998, \nat, 395A, 14}
\bibitem[Sadler, Jenkins \& Kotanyi (1989)]{1989MNRAS.240..591S} Sadler, E. 
M., Jenkins, C. R. \& Kotanyi, C. G. 1989, \mnras, 240, 591 
\bibitem[Schilizzi, et al. (1975)]{1975ApJ...201..263S} Schilizzi, R. T., 
Cohen, M. H., Romney, J. D., Shaffer, D. B., Kellermann, K. L.,
Swenson, G. W. , Jr., Yen, J. L., Rinehart, R. 1975, \apj, 201, 263
\bibitem[]{}{Slee, O. B., Sadler, E. M., Reynolds, J. E., Ekers, R. D. 1994, MNRAS, 269, 928}
\bibitem[]{}{Thompson, A. R., Clark, B. G., Wade C. M., \& Napier, P. J. 1980, \apjs, 44, 151}
\bibitem[]{}{van Dyk, S., \& Ho, L. C.~1997, IAU Coll. 164, A.~Zensus,
G.~Taylor, J.~Wrobel (eds.), ASP Conf.~Ser.~Vol.~144, p. 205}
\bibitem[Wrobel \& Heeschen (1984)]{1984ApJ...287...41W} Wrobel, J. M. \& 
Heeschen, D. S. 1984, \apj, 287, 41 
\bibitem[Wrobel \& Heeschen (1991)]{1991AJ....101..148W} Wrobel, J. M. \& 
Heeschen, D. S. 1991, \aj, 101, 148 
\bibitem[]{}{Yi, I., \& Boughn, S. P. 1998, ApJ, 499, 198}
\end{thebibliography}
\end{document}